\documentclass[sigconf,authorversion,nonacm]{acmart}
\usepackage{tikz}

\AtBeginDocument{%
  }

\setcopyright{cc}
\setcctype{by}
\copyrightyear{2026}
\acmYear{2026}
\acmDOI{}
\acmConference[SIGCSE '27]{Make sure to enter the correct
  conference title from your rights confirmation email}{February 17--20,
  2018}{Sacramento, CA}
\acmISBN{978-1-4503-XXXX-X/2018/06}
\makeatletter
\gdef\@copyrightpermission{
   \begin{minipage}{0.3\columnwidth}
     \href{https://creativecommons.org/licenses/by/4.0/}{\includegraphics[width=0.90\textwidth]{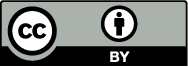}}
   \end{minipage}\hfill
   \begin{minipage}{0.7\columnwidth}
     \href{https://creativecommons.org/licenses/by/4.0/}{This work is licensed under a Creative Commons Attribution International 4.0 License.\\ \copyright 2026 Copyright held by the owner/author(s).}
   \end{minipage}
   \vspace{5pt}
}
\makeatother

\graphicspath{{fig/}}

\newcommand{\tool}{\texttt{Argus}}

\makeatletter
\def\ACM@cc@type{by} 
\makeatother

\begin{document}

\title{\tool{}: Academic Integrity in the Era of Generative AI}





\author{David Racovan}
\affiliation{%
  \institution{Purdue University}
  \city{West Lafayette}
  \state{Indiana}
  \country{USA}}
\email{dracovan@purdue.edu}

\author{Ajay Rawat}
\affiliation{%
  \institution{Purdue University}
  \city{West Lafayette}
  \state{Indiana}
  \country{USA}}
\email{rawat10@purdue.edu}

\author{Christopher K. May}
\affiliation{%
  \institution{Purdue University}
  \city{West Lafayette}
  \state{Indiana}
  \country{USA}}
\email{may5@purdue.edu}

\author{Jeffrey A. Turkstra}
\affiliation{%
  \institution{Purdue University}
  \city{West Lafayette}
  \state{Indiana}
  \country{USA}}
\email{jeff@cs.purdue.edu}

\renewcommand{\shortauthors}{Racovan and Turkstra, et al.}

\begin{abstract}
    The rapid proliferation of large language models (LLMs) in the context of education has introduced significant challenges in enforcement of academic integrity, especially in programming courses. We present \tool{}, an automated detection system for LLM-assisted student work in undergraduate C programming assignments. \tool{} integrates behavioral and stylistic indicators to create a holistic picture of the student’s progress through an assignment and surfaces anomalies that point to potential misuse of LLM assistance. We quantify and analyze data over six years of Spring semester offerings in a large-enrollment CS2 course at Purdue University using \tool{}, finding that 45\% of enrolled students exhibited patterns consistent with LLM-assisted code development in Spring 2026. To contextualize these results, we analyze the relationship between flagged LLM use and student performance on written, in-person proctored examinations, and find a significant negative correlation. We also explore the problem of mitigating false positives, recognizing that erroneous accusations of academic integrity carry significant consequences for students and instructors alike, particularly in the context of large enrollment courses. We argue that any automated detection system must be accompanied by a structured process for human review. We discuss the consequences for future course design, changing academic policy as these tools become more ubiquitous, and the pedagogical implications of LLM-based tools in computer science education.
\end{abstract}

\ccsdesc[500]{Applied computing~Education}
\ccsdesc[500]{Social and professional topics~Student assessment}
\ccsdesc[500]{Information systems~Near-duplicate and plagiarism detection}

\keywords{Academic integrity, AI-generated code detection, Large language models, Programming education, Software tools}


\maketitle

\section{Introduction}

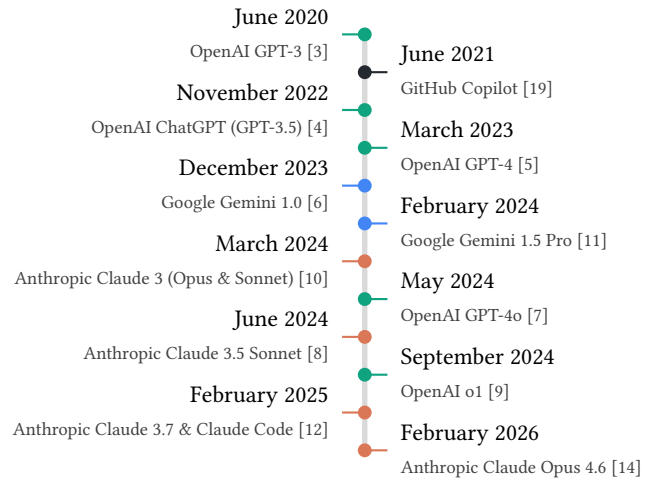
\begin{figure}[h]
\begin{tikzpicture}[very thick]

\definecolor{openaicolor}{HTML}{10a37f}
\definecolor{anthropiccolor}{HTML}{d97757}
\definecolor{googlecolor}{HTML}{4285f4}
\definecolor{githubcolor}{HTML}{24292f}
\draw[line width=2pt, gray!30] (0, 0) -- (0, -5.5);

\def\timevent#1#2#3#4#5{
    \filldraw[#4] (0, #1) circle (2pt);
    \ifnum#5=1
        \draw[#4, thick] (0, #1) -- (0.3, #1);
        \node[anchor=west, align=left] at (0.35, #1) {#2\\[0.05cm] \footnotesize\textcolor{darkgray}{#3}};
    \else
        \draw[#4, thick] (0, #1) -- (-0.3, #1);
        \node[anchor=east, align=right] at (-0.35, #1) {#2\\[0.05cm] \footnotesize\textcolor{darkgray}{#3}};
    \fi
}

\timevent{0}{June 2020}{OpenAI GPT-3 \cite{gpt3}}{openaicolor}{-1}
\timevent{-0.5}{June 2021}{GitHub Copilot \cite{github_copilot}}{githubcolor}{1}
\timevent{-1.0}{November 2022}{OpenAI ChatGPT (GPT-3.5) \cite{chatgpt}}{openaicolor}{-1}
\timevent{-1.5}{March 2023}{OpenAI GPT-4 \cite{gpt-4}}{openaicolor}{1}
\timevent{-2.0}{December 2023}{Google Gemini 1.0 \cite{gemini-1.0}}{googlecolor}{-1}
\timevent{-2.5}{February 2024}{Google Gemini 1.5 Pro \cite{gemini-1.5}}{googlecolor}{1}
\timevent{-3.0}{March 2024}{Anthropic Claude 3 (Opus \& Sonnet) \cite{claude-3}}{anthropiccolor}{-1}
\timevent{-3.5}{May 2024}{OpenAI GPT-4o \cite{gpt-4o}}{openaicolor}{1}
\timevent{-4.0}{June 2024}{Anthropic Claude 3.5 Sonnet \cite{claude-3.5}}{anthropiccolor}{-1}
\timevent{-4.5}{September 2024}{OpenAI o1 \cite{chatgpt-o1}}{openaicolor}{1}
\timevent{-5.0}{February 2025}{Anthropic Claude 3.7 \& Claude Code \cite{claude-code}}{anthropiccolor}{-1}
\timevent{-5.5}{February 2026}{Anthropic Claude Opus 4.6 \cite{claude-opus-4-6}}{anthropiccolor}{1}

\end{tikzpicture}

  \caption{Timeline of LLM releases}
  \Description{Timeline of LLM releases}
  \label{fig:timeline}
\end{figure}

Large language models (LLMs) like ChatGPT, Claude, and Gemini \cite{chatgpt,claude-3,gemini-1.0} have recently seen substantial progress and growth in their capabilities (see Figure \ref{fig:timeline}), which has introduced new challenges in computer science education. 

As these tools have become increasingly accessible, students can now generate solutions to programming assignments and debug code without needing to learn the underlying concepts themselves \cite{Jiang_Wang_Shen_Kim_Kim_2024}.

This paper presents \tool{}, a static analysis system for classifying potentially LLM-assisted code generation in an undergraduate introductory C programming course. We report on six Spring semester offerings of the course: 2020, and 2022--2026. 
All six were taught by the same instructor, and homework assignments were of a similar format. \tool{} was developed and utilized during the Spring 2026 semester, but our analysis extends to anonymized prior student data. 

\section{Related Works}
Classic approaches to plagiarism detection in student code include MOSS, which analyzes similarities in source code between students \cite{bowyer_experience_1999}. While effective at detecting instances where students reference each other’s solutions during the programming process, LLM-generated code may share no surface-level resemblance to other submissions in the class \cite{ramachandra_detecting_2026,Taylor_Blair_Glenn_Devine_2023}. This renders traditional, similarity-based approaches largely ineffective for detecting academic dishonesty. 

Recent work has explored alternative approaches to detect the use of LLMs for code generation. Some have attempted to use general text-generation models like GPTZero \cite{GPTZero}, DetectGPT \cite{mitchell_detectgpt_2023}, or Sapling \cite{Sapling} to classify code as student-written or LLM-generated. Pan et al. \cite{pan_assessing_2024} show that existing detectors perform poorly in this task, noting accuracies near 0.5. Similarly, Argotty and Manrique \cite{esteban_cuellar_argotty_ai-generated_2026} find that existing detectors tend to prioritize either recall or accuracy, meaning that detectors with a high detection rate of LLM-generated code also have a significant false positive rate. This raises concerns when utilized in an educational format, where accusations of academic misconduct can carry large consequences for students. Additionally, they have shown that minor obfuscation of LLM-generated code tends to lead to significant declines in the rate with which detectors classify LLM-generated code correctly.

Others have created models that are specialized for detecting AI-generated code, rather than being generalized on text. A model created by Xu and Sheng \cite{xu_detecting_2024} uses code’s perplexity and burstiness to determine a score, improving performance over a GPTZero baseline, though this research was limited by testing on a single code generation model, OpenAI's \texttt{text-davinci-003} \cite{text-davinci-003}. Oedingen et al. \cite{oedingen_chatgpt_2024} show similar successes but with language constraints (Python) and restricted evaluation models. Ramachandra et al. \cite{ramachandra_detecting_2026} focused on training a model within the context of a programming course where assignments do not change on a term-by-term basis, and achieved an F1 score of 0.98-0.99; this is also limited by the portability of the tool to other institutions, along with being trained on specific LLMs (Gemini and OpenAI models).

A separate method that has been proposed is the detection of style anomalies in code, which are defined by the researchers as ``coding styles that depart from a class’s style,'' like untaught language features or nonstandard spacing. Denzler et al. \cite{denzler_style_2024} found that a significant portion (44\%) of code written by ChatGPT triggered these checks, while 26\% of submissions in their most recent course offering had high style anomaly counts. Our study is most similar to this work as it focuses on similar aspects of student code, but we correlate these results to independent measures of student performance to attempt to validate whether detections have pedagogical significance.

\section{Methodology \& Tool Design}

In our C programming course, assignments are completed on a university-managed server to which students have SSH access. Each of the 13 assignments present in each offering requires students to implement C programs that are compiled and tested against a pre-built, hidden test suite. The suite is specific to each assignment; students are told what the test cases evaluate and receive granular feedback, but do not have access to the test source. Invoking \texttt{make} against the student's code compiles their code against the test suite, while also automatically committing their code to a per-student Git remote hosted on the same server, using an implementation described in \cite{Rodriguez-Rivera_Turkstra}, similar to \cite{reid_learning_2005}. Students are informed of this behavior both in class and via the \texttt{Makefile} output, though it requires no explicit action on their part. Intermediate commits between submissions are retained and form the basis of \tool{}'s behavioral analysis.

\tool{} processes each student's repository by analyzing the full commit history. For each commit, a set of static heuristic indicators is evaluated against the source code. These fall into two distinct categories:

\begin{enumerate}
    \item Advanced C topics not taught in the course curriculum, and
    \item stylistic patterns inconsistent with course conventions.
\end{enumerate}

\tool{} primarily targets C idioms and language features that do not provide value in the context of an assignment, like the use of advanced library functions before they are introduced, unnecessary dynamic memory allocation, or unneeded function modifiers. For example, the tool evaluates the presence of \texttt{inline} function declarations; while common in production codebases to optimize function call overhead (and therefore frequently produced by LLMs trained on such repositories), these optimizations are unnecessary for the scope of the course programming assignments, and are not taught formally. \tool{} also targets highly irregular comments, like those that mention LLM model names, which appear when students erroneously copy more content than intended.

Each indicator is weighted according to how much it appears in the student's code, along with a constant that indicates the level with which course staff has observed the indicator appearing in LLM-generated code. This was developed by testing frontier coding and general-purpose LLMs against previous assignments, and observing the generated output. These are added to create an aggregate heuristic score (H-score) which indicates the confidence with which a student's working history is determined to be LLM-assisted.

\tool{} also employs a dynamic normalization routine, which calculates the 10th-percentile class occurrence for each indicator $k$, subtracting this baseline from individual student counts. The penalty weight $w_k$ of an indicator is also scaled by the inverse of its class prevalence ($\rho_k$). Through this mechanism, if an LLM-favored idiom is used as part of an assignment, its diagnostic weight automatically attenuates to zero.

The temporal structure of the student's work across the assignment is also analyzed. The elapsed time between a student's earliest commit and their final submission is compared against a per-assignment threshold scaled to expected assignment complexity based on the average number of lines of code written by students for that assignment. A student who submits work within a very short window relative to the assignment's scope raises suspicion, especially when their span falls well outside of the class average.

Burst detection identifies individual commits where a large number of novel lines appear in a short interval. Specifically, commits where at least 30 new lines are introduced at a rate of 15 or more lines per minute are classified as such; this was empirically chosen based on observed student and staff behavior, and to minimize false positives. These bursts are consistent with pasting externally generated code rather than typing incrementally. Both span and burst detections are added to the H-score as components.

The resulting scores are surfaced through a web interface used by course staff. Students are ranked by H-score, and staff can inspect each flagged indicator in the context of the surrounding source code. All flagged cases undergo manual review before any academic integrity action is taken, by policy. We implement features to ensure that course staff can perform these checks quickly, especially in offerings with large enrollment. These include shortcuts to move through large amounts of code quickly, temporal charts to show code in the context of the amount of time it took to write, and charts that show how the H-score changed during the lifetime of the assignment. In many cases, a high H-score during the course of the assignment followed by a sharp decrease immediately prior to the deadline can serve as an additional factor for course staff to consider, as it usually indicates a student attempting to remove parts of their code that they may consider suspicious.

At institutions where similar infrastructure is not present, the tool may still function on the basis of static heuristic indicators. If development environments that provide insight into the student's working history with their code are available, this will provide the most accurate data; we have found that intermediate steps often reveal more indicators than a student's final submission.

\subsection{Limitations and False Positives}

Any automated system operating within the context of academic integrity must be designed with the risk of false positives as a central concern. Accusations of academic dishonesty invoke significant anxiety for students, and false convictions can have detrimental future academic and career effects. In the context of \tool{}, it is designed such that a high H-score is never treated as evidence of misconduct. Its primary purpose is to expedite the academic integrity triage process, especially in courses with large enrollment where individual, manual review of every student's code and assignment history is infeasible. 

Several benign behaviors can produce elevated scores; a student who has prior programming experience may use language patterns not taught or taught late in the course, especially if they have previously been enrolled in the course. A student who drafts code in an external editor may also produce burst indicators if they paste sections of their code between rapid commits. To mitigate this, staff take each finding in the context of surrounding code, and observe the student's full working history.

\subsection{Analysis}
\label{analysis}
To validate \tool{} as a reliable diagnostic instrument that other practitioners can confidently adopt, we compare its generated H-scores with objective measures of student performance and learning. 
Examinations are closed-note, written, and proctored, and the use of any external assistance is prohibited by course policy. This helps to serve as an independent baseline of student performance, where students are assessed on their ability to complete similar programs to those completed in prior assignments. Three in-person, closed-book exams are administered each semester. The two midterm exams involve writing substantial amounts of code drawing from concepts covered on the homework assignments while final exams are typically composed of shorter, simpler questions. H-scores are averaged across all 13 homework assignments per student, and exam scores are expressed as a percentage of total points.

For the purposes of this research, \tool{} was executed against  data from prior offerings of the course in the same way in which it was used in the Spring 2026 offering. All offerings presented in this research used similar syllabi under the same professor, with similarly-written examinations. This produced a dataset of H-scores for each student in each semester. A manual review process did not take place, as the data was not used to make individual decisions on academic integrity.

\section{Results}

\subsection{Elevated H-scores}

\begin{figure}[H]
  \centering
  \includegraphics[width=\linewidth]{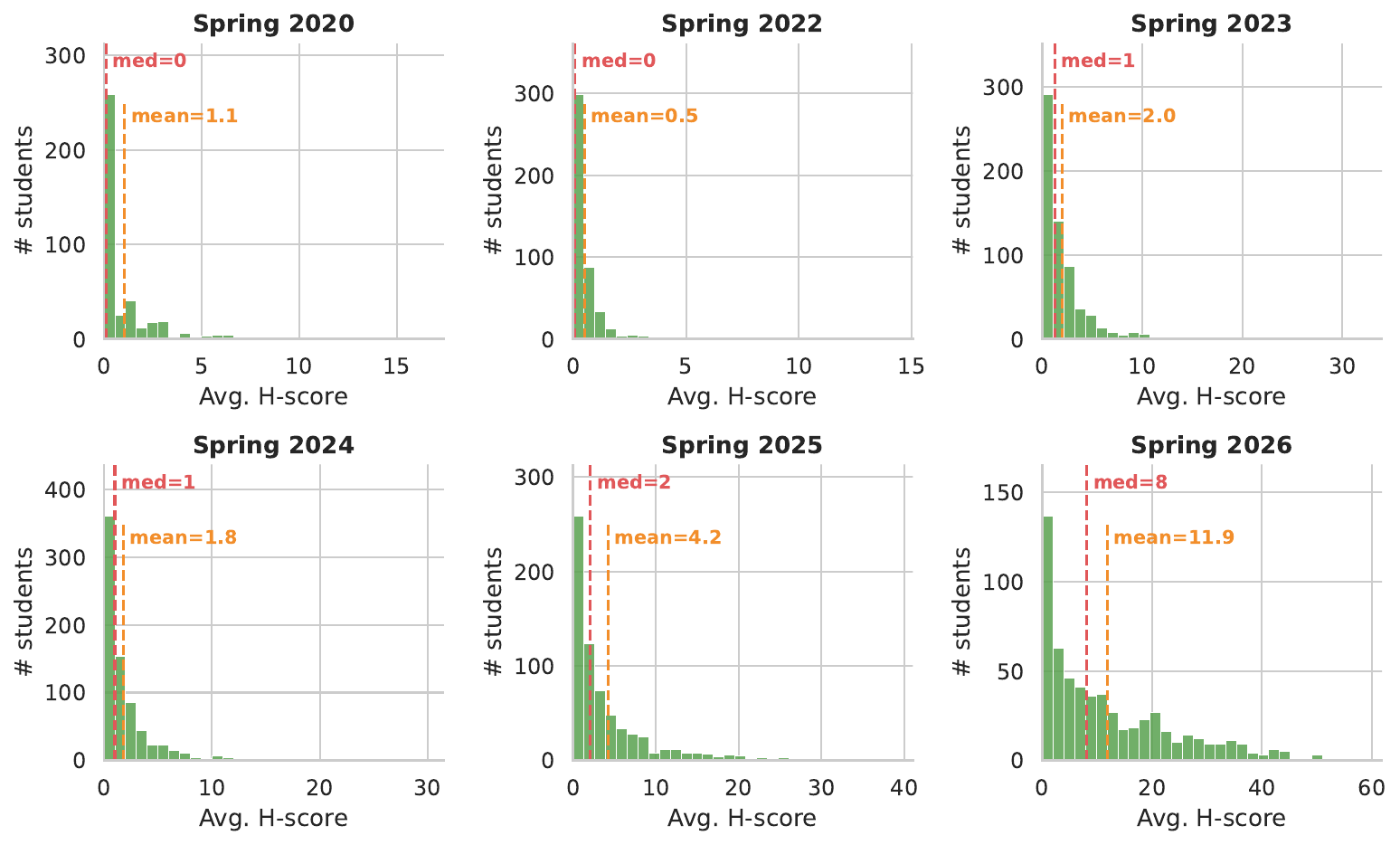}
  \caption{Distribution of Avg. H-scores, Per Semester}
  \Description{Graph of Distribution of Avg. H-scores, Per Semester}
  \label{fig:distribution}
\end{figure}

Across all six offerings of the course, \tool{} processed 3,893 distinct student-semester pairs, with a total of 42,695 individual homework submissions. Score distributions were heavily right-skewed in all semesters, as shown in Figure \ref{fig:distribution}.

\begin{figure*}[t]
  \centering
  \includegraphics[width=\linewidth]{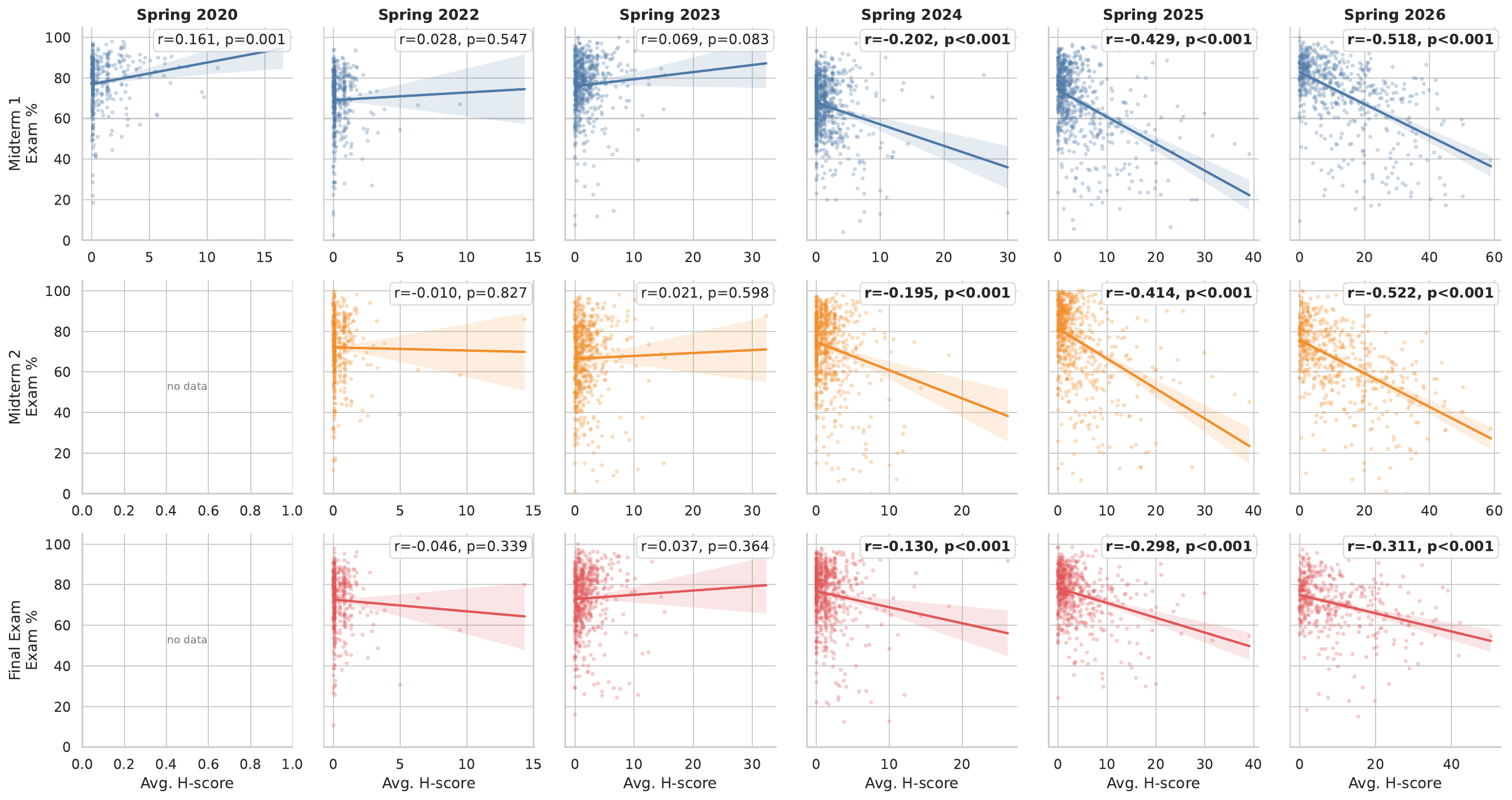}
  \caption{H-score vs. Exam Score, Per Semester; bands visualize a 95\% confidence interval of the regression fit.}
  \Description{Graph of H-score vs. Exam Score, Per Semester}
  \label{fig:scatter}
\end{figure*}

The near-zero mean scores in Spring 2020 (mean = $1.1$) and Spring 2022 (mean = $0.5$) validate \tool{}'s calibration; frontier LLM-based coding assistants were not widely available to the public during these offerings, so \tool{} accordingly found little to flag. The modest increase in Spring 2023 (mean = $2.0$) and Spring 2024 (mean = $1.8$) coincides with the initial public release and growing adoption of general-purpose LLMs, though aggregate scores remained low. This suggests that LLM-assisted development had not yet become widespread among students in the course.

A more pronounced shift appears in Spring 2025 (mean = $4.2$), and Spring 2026 (mean = $11.9$). This suggests a large change in student behavior towards the use of LLM-based assistance, rather than a change driven by a small number of outliers. Maximum observed H-scores also increased substantially, reaching 81 in Spring 2026 compared to a maximum of 40--45 in earlier offerings. These trends align with the rapid spread of coding assistants with capabilities equal to or beyond the complexity of course assignments, as charted in Figure \ref{fig:timeline}. This is consistent with the hypothesis that \tool{}'s results in earlier years demonstrate a low false-positive rate when evaluating historical, pre-LLM submissions as human-authored.

\subsection{H-scores and Exam Performance}

To assess whether the identification of LLM use by \tool{} is associated with reduced mastery of course material, this analysis compared average homework H-scores against performance on the examinations mentioned in Subsection \ref{analysis} on a per-student basis. In Spring 2020, a project was used in lieu of exams beyond Midterm 1 due to COVID-19, so the analysis for that offering is limited. All other semesters were analyzed on scores from two midterms and a final exam.

\begin{figure}[H]
  \centering
  \includegraphics[width=\linewidth]{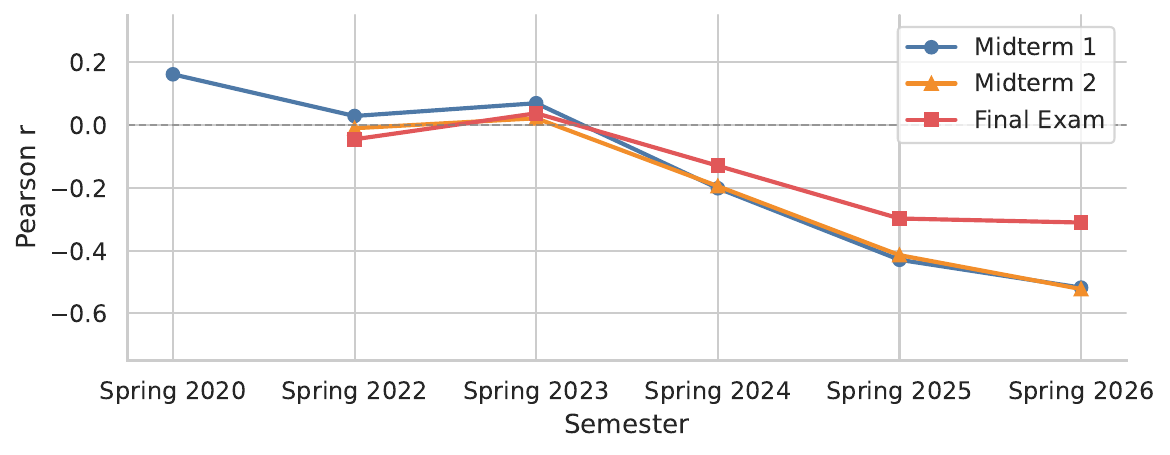}
  \caption{Correlation in H-score vs. Exam Score Over Time}
  \Description{Graph of Correlation in H-score vs. Exam Score Over Time}
  \label{fig:correlation}
\end{figure}

When analyzed on a per-semester basis, the relationship between H-score and exam performance changes markedly over time, as shown in Figures \ref{fig:scatter} and \ref{fig:correlation}. In Spring 2020, the correlation was weakly \textit{positive} ($r=0.161$), and in Spring 2022 and 2023 it was near-zero. This is consistent with the low mean H-scores in those years, since the heuristic has little variance to correlate against. Beginning in Spring 2024, a negative relationship begins to emerge ($r=-0.244$), and strengthens in Spring 2025 ($r=-0.450$) and Spring 2026 ($r=-0.537$). In Spring 2026, the per-exam breakdown has correlations of $r=-0.518$ for Midterm 1 and $r=-0.522$ for Midterm 2, with a somewhat weaker relationship on the Final Exam ($r=-0.311$). This is consistent with final exams requiring less coding overall.

\begin{figure*}[t]
  \centering
  \includegraphics[width=\linewidth]{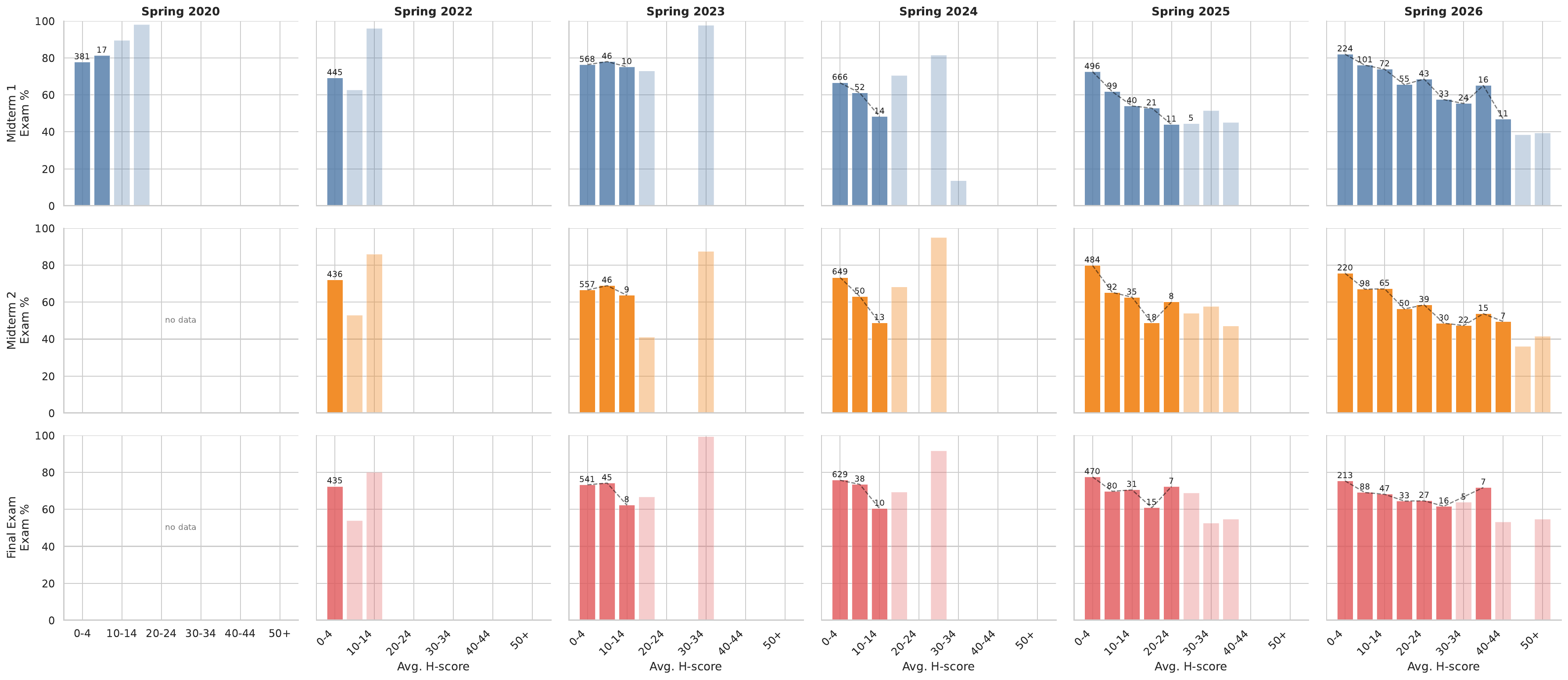}
  \caption{H-score vs. Exam Score; lighter boxes indicate groups with $n\le5$.}
    \Description{Graph of H-score vs. Exam Score, Per Semester}
    \label{fig:buckets}
\end{figure*}

The relationship between H-score and exam performance is also evident when students are grouped into scoring regions (Figure \ref{fig:buckets}). Overall, students with an average H-score in the 0--4 range ($n=2780$) achieved a mean exam percentage of 73.6\%, and performance declined steadily as H-scores increased: 68.7\% for scores in the 5--9 range, 65.8\% for 10--14, 60.0\% for 15--19, and 55.9\% for 25--29, falling to 42.5\% for scores above 50. The decline is most pronounced in later semesters, where there is sufficient score variance to populate the higher groups. Similar findings are shown by Pierce \& Zilles \cite{pierce_investigating_2017} and Chen et al. \cite{Chen_Lewis_West_Zilles_2024}, who find that plagiarism is negatively correlated with final grades in their courses.

\subsection{Withdrawal and Course Completion}

We examined whether students who withdrew from the course after Midterm 1 (students who attended Midterm 1 but neither Midterm 2 nor the Final Exam) exhibited higher H-scores than students who completed the course. This criterion was chosen to avoid drawing conclusions about students who withdrew early in the semester, before a significant portion of the course was completed. Spring 2020 is excluded, due to the availability of only a single exam.

\begin{figure}[H]
  \centering
  \includegraphics[width=\linewidth]{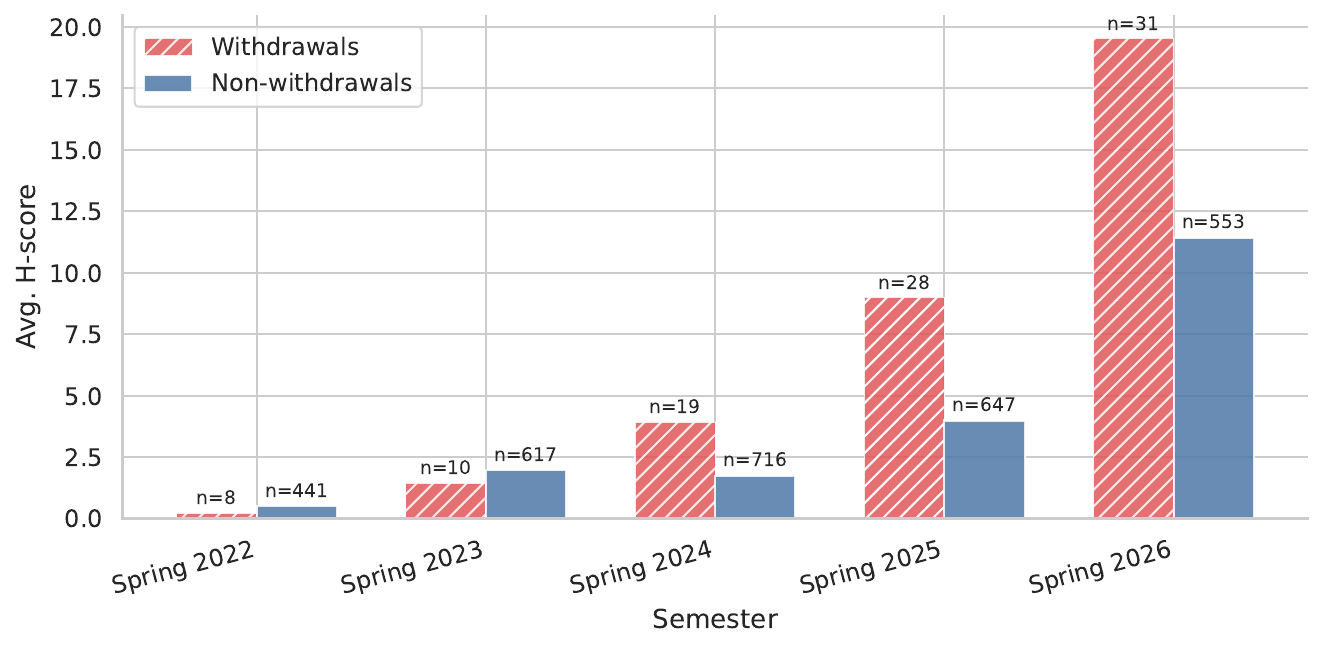}
  \caption{Mean H-score vs. Withdrawals by Semester}
  \Description{Graph of Mean H-score vs. and Withdrawals by Semester}
  \label{fig:withdrawals}
\end{figure}

Across the five remaining semesters, 96 students (approximately 3.1\% of the population) met the withdrawal criteria. Their mean H-score was 9.9, compared to 3.9 among the 2,974 students who completed the course. When evaluated per-semester, this disparity widened notably over time; these effects were the smallest in Spring 2022 and Spring 2023, but in Spring 2026, withdrawn students averaged a H-score of 19.5 against 11.4 for continuing students, detailed in Figure \ref{fig:withdrawals}. A two-sample Kolmogorov-Smirnov test indicates that the two groups are drawn from significantly different distributions ($D=0.362$, $p<0.001$). The association between elevated H-scores and course withdrawals is consistent with a scenario in which students who rely heavily on external assistance accumulate less understanding of the material over the course of the semester and subsequently struggle to apply it in a proctored context.

\subsection{Case Study: Spring 2026}

The results presented rely on \tool{}'s H-scores as the sole proxy for LLM-assisted development. While this enables longitudinal comparison across all six offerings, human review remains necessary to draw conclusions for an individual student. In Spring 2026, we implemented a structured manual review process, in which course staff examined students with high H-scores as identified by \tool{}. Students were ranked on H-score both cumulatively and per-assignment and reviewed top-to-bottom. Course staff made a binary determination of whether the evidence was sufficient to pursue the case, per course policy. Three separate members of course staff evaluated students independently, then met to reach consensus. Extreme benefit of the doubt was given throughout to avoid false positives; students were only determined to have used LLM-assisted tools if the evidence was remarkably strong. 

To characterize the practical cost of review, we tracked the time it required. Reviewers examined the top 300 students by H-score, plus any student in the top 150 for an individual assignment. Three course staff spent approximately 8 hours each, with a median review time of roughly 1.5 minutes per student: clearly suspicious commits were often identified in under a minute, while borderline cases required 5 minutes or more of examining the full working history. The temporal charts and per-indicator navigation were essential to this throughput; in informal comparison, reviewing a raw repository without the tool took upwards of 10 minutes per student. 

Among the top 150 students by H-score, we noted 4 instances where there was insufficient evidence to escalate their cases. In these, the indicators had a benign explanation in context: bursts traced to code and comments copied from the assignment description itself, or untaught language features were used coherently throughout, consistent with prior experience. 

\begin{figure}[h]
  \centering
  \includegraphics[width=\linewidth]{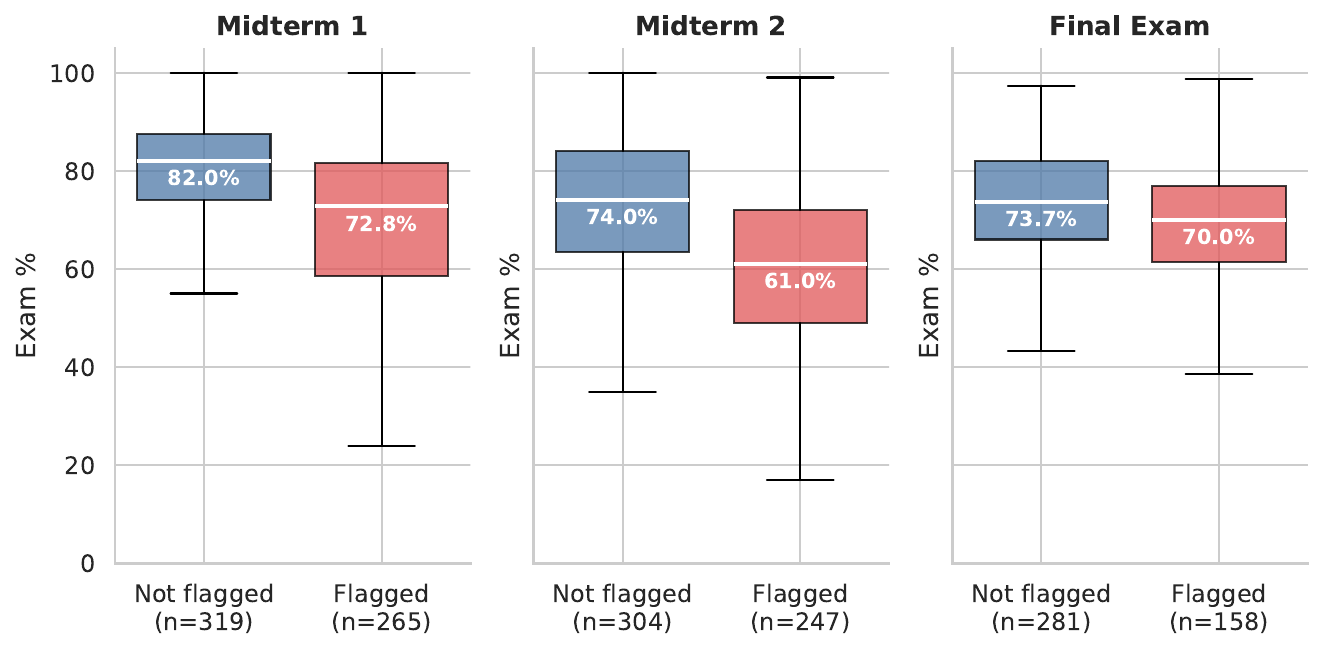}
  \caption{Exam Score, by Flagged vs. Not Flagged (Spring 2026)}
  \Description{Graph of Exam Score, by Flagged vs. Not Flagged (Spring 2026)}
  \label{fig:flagged_box}
\end{figure}

Of the 584 students who completed the first midterm exam in the Spring 2026 offering, 267 (45.7\%) were flagged following manual review, and 317 were not. The prevalence of flagged cases in this offering is consistent with the substantially increased H-scores observed in Spring 2026 relative to other semesters. Flagged students performed notably worse on all three examinations than their non-flagged peers (Figure \ref{fig:flagged_box}). Averaged across all three exams, flagged students achieved a mean of 61.9\% (median 65.9\%), compared to 73.5\% (median 75.7\%) for non-flagged students. The gap was largest on Midterm 2, where flagged students averaged 59.0\% versus 72.2\% for non-flagged students.

Among flagged students, performance declined as the number of flagged assignments increased, also suggesting a relationship between amount of LLM-based assistance and exam outcomes. Students with a single flagged assignment averaged 69.4\%, close to the non-flagged mean of 73.5\%. Performance then declined progressively; students with 2 flagged assignments averaged 65.0\%, students with 5 averaged 59.0\%, students with 6 averaged 53.5\%, and those with 8 or more averaged 54.4\%. 

We also utilized MOSS in parallel with \tool{} for the Spring 2026 offering, but detected cases were significantly decreased compared to prior offerings, with an average of about 2-3 suspicious student-to-student plagiarism cases per assignment. Since LLM-assisted submissions often do not bear similarities that tools like MOSS depend on, dishonest submissions become progressively less visible to similarity-based tools.

The large performance gap between flagged and non-flagged students on in-person, proctored examinations supports the interpretation that flagged students had not developed the independent mastery of course material that the homework assignments were designed to build. The gap widening as the number of assignments flagged increases further suggests that the amount of LLM reliance, not merely its presence, is predictive of reduced outcomes.

\section{Discussion \& Reflections}

\subsection{Automated Detection and Human Review}

The scale of apparent LLM adoption in Spring 2026 surfaces a tension any institution deploying detection tooling will face: the volume of cases can quickly exceed what course staff can review, while the consequences of acting on an automated score alone are too serious to accept. Ideally, humans would review all code from every student, but this is infeasible at scale; Spring 2026 alone produced 381,748 individual commits across 7,769 submissions. \tool{} was designed to help resolve this tension rather than decide student outcomes. By ranking students on H-score and surfacing the specific indicators behind it, staff can move through a large case queue efficiently. We accept a non-zero false negative rate in doing so, but by surfacing the students harming their learning the most, action can be taken early to correct these behaviors.

The failure mode of any similar detection tool used in the context of academic integrity is asymmetric; a missed case of LLM use affects only that student's own learning, but a false accusation can have lasting consequences for a student's academic record and wellbeing. We argue that the appropriate standard for any detection system is not whether it can identify misconduct autonomously, but whether it can make the human review process fast and structured enough to be feasible for large-enrollment courses.

\subsection{The Scale of LLM Adoption}

The most striking finding of this research is the speed with which the presented correlation increased over time. In Spring 2020 and Spring 2022, \tool{} flagged very little, and the relationship between its scores and exam outcomes were near-zero. Within four years, 45\% of the students in the same course, taught by the same instructor, with structurally similar assignments, exhibited patterns consistent with LLM-assisted development. This suggests that the question facing CS2 courses is no longer whether students will have access to and use capable coding assistants, but how course design and policy should respond to the assumption that they do.

\subsection{Changing Academic Policy}

The findings also raise questions about how academic integrity policy should evolve as LLM-based tools become more capable and more ubiquitous. The definition of unauthorized assistance continues to become more nebulous; there is a pedagogical difference between a student using an LLM to debug a single function versus generating an entire submission, but this distinction may not be reflected in common institutional policies. However, any new guidelines must balance this nuance with the practical requirement that policies remain straightforward to understand and enforce. As detection tools improve, institutions will need policies that are both adaptable to new technologies and highly unambiguous in practice.

\section{Conclusion}

We have presented \tool{}, a static analysis system for detecting patterns consistent with LLM-assisted code development in an undergraduate C programming course, applied across six semester offerings from Spring 2020 through Spring 2026. H-scores were near-zero in the pre-LLM era and rose substantially as LLM-based coding assistants became publicly available, validating that the tool reflects real changes in student behavior. Average H-score was negatively correlated with performance on in-person, closed-note examinations, with correlations strengthening in more recent offerings. The pedagogical implication is that students who rely heavily on LLM assistance do not develop the independent programming ability the course is designed to produce.

We emphasize that \tool{} is a triage tool, not designed to render automated determinations of academic dishonesty; institutions deploying automated detection should invest equally in the review process around it. Given the pace of LLM capability growth, the patterns observed in Spring 2026 will likely grow more prevalent. We hope \tool{} and the methodologies described here offer a useful foundation for institutions responding to this shift, and that our longitudinal analysis provides a strong baseline against which future detection methods can be measured.

\clearpage
\bibliographystyle{ACM-Reference-Format}
\bibliography{bibliography}

\end{document}